\documentclass[10pt,twosided]{article}
\date{}
\usepackage{amsmath}
\usepackage{color}
\usepackage{amsfonts}
\usepackage{verbatim}
\usepackage{amssymb}
\usepackage{bbold}
\usepackage{graphicx}
\numberwithin{equation}{section}

\begin{document}
\title{Our Trysts with `Bal' and Noncommutative Geometry}
\author{Biswajit Chakraborty$^{1}$\footnote{dhrubashillong@gmail.com}, Partha Nandi$^2$, Sayan Kumar Pal$^2$, Anwesha Chakraborty$^2$}
\date{%
    $^1$Department of Physics, School of Mathematical Sciences,\\
    Ramakrishna Mission Vivekananda Educational and Research Institute\\
    Belur Math, Howrah-711202, India\\%
    $^2$S N Bose National Centre or Basic Sciences,\\
    Block-JD, Sector-III, Saltlake - 700106, India\\[2ex]%
   \textit{ In Honor of A.P.Balachandran on the Occasion of His 85th Birthday \\(To be published in the Festschrift volume: Particles, Fields and Topology by World Scientific, Singapore)}.
}

\maketitle
	\begin{abstract}
	This contributory article begins with our fond and sincere reminiscences about our beloved Prof. A.P. Balachandran. In the main part, we discuss our recent formulation of quantum mechanics on (1+1)D noncommutative space-time using Hilbert-Schmidt operators. As an application, we demonstrate how geometrical phase in a system of time-dependent forced harmonic oscillator living in the Moyal space-time can emerge.
	\end{abstract}
\section{ Down the Memory Lane with Prof. A.P. Balachandran}

It is indeed a great privilege for us to get the opportunity to dedicate this note to one of our role models and teacher - a doyen of theoretical physics: Professor A.P. Balachandran, on his 85th birthday. To start with, let us begin with some brief nostalgic reminiscences of the past. One of us, B.C. knows Professor A.P. Balachandran, or simply Bal as we fondly refer to him, for almost four decades now. Although, I never had the privilege of co-authoring any paper with him, I share with him our common alma-mater; we both are alumni of IMSc (Institute of Mathematical Sciences, Chennai) where we have carried out our Ph.D. works, a quarter century apart.  I had joined IMSc in 1985, soon after E.C.G. Sudarshan took charge of this institute as its director and Prof. C.S. Seshadri, R. Balasubhramanian joined the Mathematics department and IMSc started getting very distinguished visitors like Bal in a regular basis.  It was this rejuvenated academic environment that stimulated our passion for pursuing high quality research work to the best of our individual capacity. Importantly, Bal used to visit IMSc at least for once/twice every year to collaborate and give seminars/talks  on new and interesting developments happening in theoretical Physics. He also used to deliver some pedagogical lecture series for the graduate students in IMSc. And once we got ourselves familiarized with the basics of mathematical aspects like topology, differential geometry, group theory  etc. his lectures and/or papers started becoming accessible and I was fascinated by the beauty and depth of the whole approach and eventually he became a role model for me, both as a researcher and a teacher.\\
Among the many areas that he has contributed, I would, particularly, like to mention about his enormous and deep contribution for understanding the symmetry aspects of noncommutative spaces, as captured by quantum groups/Hopf algebras and he has also written several books on the subject \cite{books}. In fact, in one of his colloquia, delivered in the Saha Institute of Nuclear Physics, Kolkata, in early 2000, he gave an introductory lecture on noncommutative geometry- a new emerging area at that time. Particularly, his elaborated explanation on Gelfand-Naimark theorem was absolutely fascinating to me and in retrospect I feel that this single event played a pivotal role in triggering my persistent fascination towards this subject - an area of my current interest.

On the other hand, the other co-authors have got to know Bal personally only since 2018, when he participated as a keynote speaker in the international conference on noncommutative geometry held at S.N. Bose Centre, Kolkata. Although we were familiar with his works which served as our references and check-points for our previous papers, this was the first occasion that we got to meet him personally and carry out intense academic discussions. We were quite enthralled to see his passion for physics even at this ripe age and patience to answer our naive questions. Subsequently, we have maintained our regular academic contacts in the online mode where he makes himself always available for physics discussions. Besides, almost all of our recent papers were written after incorporating all of his deep suggestions. Finally, he was quite generous to give all of us (P.N. , S.K.P. , A.C.) the opportunity to speak at his famous Room-316 meetings from Syracuse University, which are being re-organised these days in online mode after the pandemic got started.

\section{Noncommutative Geometry: Quantum space-time}
The celebrated idea of noncommutativity in modern physics has attracted a lot of interest theoretically in particle physics and condensed matter physics. Additionally, noncommutativity has played an important role in the study of quantum gravity at the Planck scale over the previous two decades. To motivate the emergence of noncommutative space-time or more generally quantum space-time, one can think of a superposition of two mass distributions. As has been argued by Penrose in \cite{Roger} that as a feedback through Einstein's GR, this will give rise to a superposed geometry/space-time. Now, such a quantum space-time is likely to loose its time-translational symmetry resulting in the uncertainty of energy $\delta E$ and time $\delta t$,  indicating a finite lifetime $\sim \frac{\hbar}{\delta E}$ of the system. This heuristic argument indicates that one needs to reformulate the quantum theory without classical time, rather time should be promoted as an operator-valued coordinate, along with other operator-valued spatial coordinates (see also \cite{fred,Dop}). And we mention, in this context, the status of time in quantum gravity is an age-old problem. In fact, its status in QM itself is a bit ambiguous. One can, in fact, recall Pauli's objection in this context \cite{pauli} and this ambiguity can result in other allied problems (for example see \cite{pull}). And the physics of such quantum systems should have its manifestations, atleast in principle, even in sub-Planckian energy scales which also can even be effectively non-relativistic, like non-classical features in primordial gravitational waves \cite{soda}. Again, here we can cite many of the seminal works of Bal in this direction \cite{Bal2}. In fact, we were motivated primarily through these papers to pursue such studies and take it forward, as we describe in the sequel.

In the present article, we discuss about our recent investigations and formulation of QM in (1+1)D noncommutative (NC) space-time \cite{partha, partha2}. Thereafter, as an application, we discuss about the possibility of obtaining emergent Berry phase in a system inhabiting NC spacetime. We start by giving a brief outline of the Hilbert-Schmidt operator based formulation of noncommutative quantum mechanics (NCQM) which will serve as the basic mathematical framework for the present treatment.
\section{Space-time NCQM using Hilbert-Schmidt operator formulation}
Although initially the Hilbert-Schmidt (HS) operator based formulation of noncommutative quantum mechanics was systematically devised, following \cite{nair}, to formulate quantum mechanics on spatial 2D noncommutative Moyal plane \cite{biswa,gauba}, here in the present article we provide a brief review of our recent works on the formulation of QM on noncommutative (1+1)D Moyal space-time (based on the results in \cite{partha, partha2}) and how the HS operator formulation can be adapted to formulate or rather used to extract an effective, consistent and commutative quantum-mechanical theory. Now, before considering the quantum theory, let us first discuss the appearance of noncommuting nature of space-time brackets just at the classical level itself. To this end, let us consider the following Lagrangian of a non-relativistic particle in (1+1)D written in the first-order form \cite{Deri}~:
\begin{equation}
	L^{\tau,\theta}= p_{\mu}\dot{x}^{\mu}+\frac{\theta}{2}\epsilon^{\mu\nu}p_{\mu}\dot{p}_{\nu}-\sigma(\tau)(p_t+H),\,\,\,\,\,\mu ,\nu = 0,1\label{e73}
\end{equation}
where $x^{\mu}=(t,x)$ and $p_{\mu}=(p_t,p_x)$ are both counted as configuration space variables. The evolution parameter $\tau$ is chosen a bit arbitrarily, except that it should be taken as a monotonically increasing function of time $`t$'. The over-head dots indicate $\tau-$ derivatives. The middle term is analogous to Chern-Simons term with $\theta$ being the corresponding parameter.
On carrying out Dirac's analysis of constraints, one arrives at the following Dirac brackets between the phase space variables.
\begin{equation}
	\{x^{\mu},x^{\nu}\}_D=\theta\epsilon^{\mu\nu};\,\,\,\,\,\{p_{\mu},p_{\nu}\}_D=0;\,\,\,\,\,\,\{x^{\mu},p_{\nu}\}_D=\delta^{\mu}\,_{\nu}\label{lev}
\end{equation}
Finally, the Lagrange multiplier $\sigma(\tau)$ enforces the following first class constraint in the system,
\begin{equation}
	\Sigma= p_t+H \approx 0 \label{e87}
\end{equation}
and can be shown to generate the $\tau$ evolution of the system in the form of gauge transformation of the theory.\\
We now elevate the Dirac brackets in (\ref{lev}) to the level of noncommuting operators in order to initiate the quantum mechanical analysis for this (1+1)D non-relativistic quantum mechanical system in presence of the space-time noncommutativity of Moyal type:
\begin{equation}
	[\hat{t},\hat{x}]=i\theta\label{e1} ~;~~~~~{\rm where,~}\theta~{\rm is~the~ NC ~parameter}
\end{equation} 
\subsection{HS operator formulation}A representation of NC coordinate algebra (\ref{e1}) can be readily found to be furnished by the Hilbert space,
\begin{equation}
	\mathcal{H}_c=Span \left\{|n\rangle = \frac{(b^{\dagger})^n}{\sqrt{n!}}|0\rangle;\,\,b|0\rangle =\frac{\hat{t}+i\hat{x}}{\sqrt{2\theta}}|0\rangle =0 \right\}\label{e3}
\end{equation} 
We now introduce the associative NC operator algebra ($\hat{\mathcal{A}}_{\theta}$) generated by ($\hat{t},\hat{x}$) or equivalently by  ($\hat{b},\hat{b}^{\dagger}$)  acting on this configuration space $\mathcal{H}_c$ (\ref{e3}) as
\begin{equation}
	\hat{\mathcal{A}}_{\theta}=\{|\psi)=\psi (\hat{t},\hat{x})=\psi(\hat{b},\hat{b}^{\dagger})= \sum_{m,n} c_{n,m}|m\rangle\langle n|\}\label{e59}
\end{equation}
which is the set of all polynomials in the quotient algebra ($\hat{\mathcal{A}}/\mathcal{N}$) in $(\hat{t},\hat{x})$ or equivalently in ($\hat{b},\hat{b}^{\dagger}$), subject to the identification of $[\hat{b},\hat{b}^{\dagger}]=1$. Thus, $\hat{\mathcal{A}}_{\theta}=\hat{\mathcal{A}}/\mathcal{N}$ is essentially identified as the universal enveloping algebra corresponding to (\ref{e1}), where $\hat{\mathcal{A}}$ is the free algebra generated by ($\hat{t},\hat{x}$) and $\mathcal{N}$ is the ideal generated by (\ref{e1}). This $\hat{\mathcal{A}}_{\theta}$ is not equipped with any inner product at this stage.\\ 
We can now introduce a subspace $\mathcal{H}_q \subset \mathcal{B}(\mathcal{H}_c) \subset \hat{\mathcal{A}}_{\theta}$ as the space of `HS' operators, which are bounded and compact operators with finite HS norm $||.||_{HS}$, which acts on $\mathcal{H}_c$ (\ref{e3}), and is given by,
\small
\begin{equation}
	\mathcal{H}_q= \left\{\psi(\hat{t},\hat{x})\equiv \Big|\psi(\hat{t},\hat{x})\Big)\in \mathcal{B}(\mathcal{H}_c) ;\, \, ||\psi||_{HS}:=\sqrt{tr_c(\psi^{\dagger}\psi)} < \infty\right\} \subset \hat{\mathcal{A}}_{\theta}\label{e4}
\end{equation}
\normalsize
where $tr_c$ denotes trace over $\mathcal{
	H}_c$ and $\mathcal{B}(\mathcal{H}_{c})\subset \hat{\mathcal{A}}_{\theta}$ is a set of bounded operators on $\mathcal{H}_{c}$.  This space can be equipped with the inner product
\begin{equation}
	\Big(\psi(\hat{t},\hat{x}),\phi(\hat{t},\hat{x})\Big):=tr_{c}\Big(\psi^{\dagger}(\hat{t},\hat{x})\phi(\hat{t},\hat{x})\Big)
	\label{iop}
\end{equation}
and therefore has the structure of  a Hilbert space on its own.  Note we denote the elements of $\mathcal{H}_c$ and $\hat{\mathcal{A}}_{\theta}$  by the angular ket $| .\rangle$ and round ket $| . )$ respectively. We now define the quantum space-time coordinates ($\hat{T},\hat{X}$) (which can be regarded as the representation of ($\hat{t}, \hat{x}$), and must be distinguished as their domain of actions are different i.e. while $(\hat{T},\hat{X})$ act on $\mathcal{H}_{q}$,~$(\hat{t},\hat{x})$ act on $\mathcal{H}_{c}$) as well as the corresponding momenta ($\hat{P_{t}},\hat{P}_{x}$)  by their actions on a state vector $|\psi(\hat{t},\hat{x}))$ $\in \mathcal{H}_{q}$ as,\\
\begin{align}
	&\hat{T}\Big|\psi(\hat{t},\hat{x})\Big)=\Big|\hat{t}\psi(\hat{t},\hat{x})\Big),\,\,\,\,\,\hat{X}\Big|\psi(\hat{t},\hat{x})\Big)=\Big|\hat{x}\psi(\hat{t},\hat{x})\Big),\nonumber\\
	&\hat{P}_x\Big|\psi(\hat{t},\hat{x})\Big)=-\frac{1}{\theta}\Big|[\hat{t},\psi(\hat{t},\hat{x})]\Big),\,\,\,\hat{P}_t\Big|\psi(\hat{t},\hat{x})\Big)=\frac{1}{\theta}\Big|[\hat{x},\psi(\hat{t},\hat{x})]\Big) \label{e5}
\end{align}
Thus, the momenta ($\hat{P_t},\hat{P_x}$) act adjointly and their actions are only defined in  $\mathcal{H}_{q}$ and not  $\mathcal{H}_c$. It may be easily verified now that, 
\begin{equation}
	[\hat{T},\hat{X}]=i\theta,~ [\hat{T},\hat{P_t}]=i\theta=[\hat{X},\hat{P_x}],~[\hat{P_t},\hat{P_x}]=0
\end{equation}
which represents the total NC Heisenberg algebra, quantum version of the classical algebra (\ref{lev}). Note, here we are working in the natural unit $\hbar= 1$.
\subsubsection{Schr\"odinger equation and an induced inner product}
As, $\theta \neq 0$, simultaneous space-time eigenstate $|x,t\rangle $ does not exist. Nevertheless, an effective commutative theory can be constructed by making use of maximally localized  events i.e. the Sudarshan-Glauber coherent states -
\begin{equation}
	|z\rangle = e^{-\bar{z}\tilde{b}+z\tilde{b}^{\dagger}}|0\rangle\,\in\mathcal{H}_c~;~~\tilde{b}|z\rangle=z|z\rangle\label{e43}
\end{equation} 
where $z$ is a dimensionless complex number and is given by,
\begin{equation}
	z=\frac{t+ix}{\sqrt{2\theta}};\,\,\,\,\,\,\,\,\,t=\langle z|\hat{t}|z\rangle , x=\langle z|\hat{x}|z\rangle\label{A4}
\end{equation}
Here $t$ and $x$ are effective commutative coordinate variables. We can now construct the counterpart of coherent state basis in $\mathcal{H}_q$ (\ref{e4}), made out of the bases $|z\rangle \equiv |x,t\rangle$ (\ref{e43}), by taking their outer product as
\begin{equation}
	|z,\bar{z})\equiv |z)=|z\rangle\langle z|=\sqrt{2\pi \theta}\,\,|x,t)\,\in \mathcal{H}_q \,\,\,\,\,\,\textrm{fulfilling}\,\,\,B|z)=z|z)\label{A8}
\end{equation}
where the annihilation operator $\hat{B}=\frac{\hat{T}+i\hat{X}}{\sqrt{2\theta}}$ is a representation of the operator $\tilde{b}$ in $\mathcal{H}_q$ (\ref{e4}).
It can also be checked that the basis $|z,\bar{z})\equiv |z)$ satisfies the over-completeness property:
\begin{equation}
	\int \,\,\frac{d^2z}{\pi}|z,\bar{z})\,\star_{V}\,(z,\bar{z}|= \int dtdx \, |x,t) \star_{V} (x,t| = \textbf{1}_q,
	\label{vnc}
\end{equation} \\
where $*_V$ represents the Voros star product and is given by,
\begin{equation}
	\star_V=e^{\overleftarrow{\partial_z}\overrightarrow{\partial_{\bar{z}}}}=e^{\frac{i\theta}{2}(-i\delta_{ij}+\epsilon_{ij})\overleftarrow{\partial_i}\overrightarrow{\partial_j}};\,\,\,\,i,j=0,1;\,\,\,x^0=t, x^1=x
\end{equation}
Then the coherent state representation or the symbol of an abstract state $\psi(\hat{t},\hat{x})$ gives the usual coordinate representation of a state just like ordinary QM:
\small
\begin{equation}
	\psi(x,t)=\frac{1}{\sqrt{2\pi\theta}}\Big(z,\bar{z}\Big|\psi(\hat{x},\hat{t})\Big)=\frac{1}{\sqrt{2\pi\theta}}tr_{c}\Big[|z\rangle\langle z|\psi(\hat{x},\hat{t})\Big]=\frac{1}{\sqrt{2\pi\theta}}\langle z|\psi(\hat{x},\hat{t})|z\rangle \label{e45}
\end{equation}
\normalsize
The corresponding representation of a composite operator say $\psi(\hat{x},\hat{t}) \phi(\hat{x},\hat{t})$ is  given by,
\begin{equation}
	\Big(z\Big|\psi(\hat{x},\hat{t})\phi(\hat{x},\hat{t})\Big)=\Big(z\Big|\psi(\hat{x},\hat{t})\Big) \,\star_V\,\Big(z\Big|\phi(\hat{x},\hat{t})\Big).
	\label{como}
\end{equation}
This establishes an isomorphism between the space of HS operators $\mathcal{H}_q$ and the space of their respective symbols. Using (\ref{vnc})  the overlap of two arbitrary states ($|\psi),|\phi)$) in the quantum Hilbert space $\mathcal{H}_q$ can be written in the form
\begin{equation} 
	(\psi|\phi) = \int dtdx ~ \psi^\ast(x,t) \star_{V} \phi(x,t)\label{innpro_voros}
\end{equation}
Therefore, to each element $|\psi(\hat{x},\hat{t})) \in \mathcal{H}_q$, the corresponding symbol $\psi(x,t) \in L_{\star}^2(\mathbb{R}^2)$, where the $*$-occurring in the subscript is a reminder to the fact that the corresponding norm has to be computed by employing the Voros star product. In order to obtain an effective commutative Schr\"odinger equation in coordinate space, we will introduce coordinate  representation of the phase space operators. To begin with, note that the coherent state representation of the actions of space-time operators  $\{\hat{X},\hat{T}\}$, acting on $|\psi)$ can be written by using (\ref{como}) as,
\begin{equation}
	\Big( x,t\Big|\hat{X} \, \psi(\hat{x},\hat{t})\Big) =\frac{1}{\sqrt{2\pi\theta}}\Big(z,\bar{z}\Big|\hat{x}\psi\Big) = \frac{1}{\sqrt{2\pi\theta}} \, \left\langle z|\hat{x}|z\right\rangle \star_{V} (z,\bar{z}|\psi(\hat{x},\hat{t}))
\end{equation}
Finally on using (\ref{e45}), we have
\small
\begin{equation}
	\Big( x,t\Big|\hat{X} \psi(\hat{x},\hat{t})\Big) = X_\theta \, \Big(x,t\Big|\psi(\hat{x},\hat{t})\Big) = X_\theta \, \psi(x,t)~;~X_\theta = x+\frac{\theta}{2}(\partial_x-i\partial_t)\label{47}
\end{equation}
\normalsize
Proceeding exactly in the same way, we obtain the representation of $\hat{T}$ as
\begin{equation}
	T_\theta = t+\frac{\theta}{2}(\partial_t+i\partial_x),\label{48}
\end{equation}
so that $[T_{\theta},X_{\theta}]=i\theta$ is trivially  satisfied. It is now trivial to prove the self-adjointness property of both $X_\theta$ and $T_\theta$, w.r.t. the inner product (\ref{innpro_voros}) in $\mathcal{H}_{q}$ by considering an arbitrary pair of different states $|\psi_1)$, $|\psi_2) \in \mathcal{H}_q$ and their associated symbols, just by exploiting associativity of Voros star product. Since momenta operators act adjointly, their coherent state representations are,
\begin{equation}
	\Big(x,t\Big|\hat{P}_t \psi(\hat{x},\hat{t})\Big)= -i \partial_t\psi(x,t)~;~~ \Big(x,t\Big|\hat{P}_x \psi(\hat{x},\hat{t})\Big)=-i\partial_x\psi(x,t)
	\label{25}
\end{equation}
The effective commutative Schr\"odinger equation in NC space-time is then obtained by imposing the condition that the physical states $|\psi_{phy})=\psi_{phy}(\hat{x},\hat{t})$ are annihilated by the operatorial version of (\ref{e87}):
\begin{equation}
	(\hat{P}_t+\hat{H})|\psi_{phy})=0;\qquad\psi_{phy}(\hat{x},\hat{t})\in \mathcal{H}_{phy}\subset \hat{\mathcal{A}}_{\theta} \label{e74}
\end{equation} 
where $\hat{H}=\frac{\hat{P}_x^2}{2m}+V(\hat{X},\hat{T})$. 
We are now ready to write down the effective commutative time dependent Schr\"{o}dinger equation in quantum space-time by taking the representation of (\ref{e74}) in $|x,t)$ basis.
Using (\ref{47},\ref{48},\ref{25}) we finally get,
\begin{equation}
	i\partial_t \psi_{phy}(x,t)= \left[-\frac{1}{2m}\partial_x^2+ V(x,t)\, \star_{V}\right] \psi_{phy}(x,t)\label{e70}
\end{equation}
One can now obtain the continuity equation as,
\begin{equation}
	\partial_t \rho_{\theta} +\partial_x J_{\theta}^{x}=0\label{e72}
\end{equation}
where 
\small
\begin{align}
	\rho_{\theta}(x,t)= \psi_{phy}^*(x,t)\,\star_{V}\,\psi_{phy}(x,t)>0\nonumber;~J_{\theta}^{x}= \frac{1}{m}\mathfrak{Im}\bigg(\psi_{phy}^*\star_{V}(\partial_x \psi_{phy})\bigg) \label{prob}
\end{align}
\normalsize The positive definite property of $\rho_{\theta}(x,t)$ indicates that it can be interpreted as the probability density at point $x$ at time $t$. However, note that for a consistent QM formulation, we ought to have $\psi_{phy}(x,t)$ to be \textquotedblleft{well}-behaved" in the sense that it should be square integrable at a constant time slice:
\begin{equation}
	\langle \psi_{phy} |\psi_{phy} \rangle_{*\,t} =	\int_{-\infty}^{\infty} \,dx\,\, \psi_{phy}^*(x,t)\star_{V}\psi_{phy}(x,t) < \infty,
	\label{op}
\end{equation}
so that $\psi_{phy}(x,t)\in L_{\star}^2(\mathbb{R}^1)$  which is naturally distinct from $L_{\star}^2(\mathbb{R}^2)$. Equivalently , at the operator level, $\psi_{phy}(\hat{x},\hat{t})$ should belong to a suitable subspace of $\hat{\mathcal{A}}_{\theta}$ (\ref{e59}) which is distinct from $\mathcal{H}_q$, as the associated symbol for the latter is obtained from inner product defined for $L_{\star}^2(\mathbb{R}^2)$ (\ref{innpro_voros}). This is the main point of departure from the standard HS operator formulation of NCQM in (1+2)D Moyal plane with only spatial noncommutativity where time is treated as a c-parameter and one works with $\mathcal{H}_q$ or equivalently with a Hilbert space $L_{\star}^2(\mathbb{R}^2)$ for the corresponding symbols \cite{biswa,gauba}.
\section{Emergence of Berry phase in a time-dependent NC quantum system }
A time-independent system like a harmonic oscillator has been shown to exhibit no modifications in the spectrum in a noncommutative space-time \cite{partha}. Therefore, the forced harmonic oscillator, having explicit time-dependence, is a good prototype system to study and look for any possible emergent geometric phases, as one of possible signals of space-time noncommutativity.
We therefore take the up the Hamiltonian of the forced harmonic oscillator in the following hermitian form for carrying out our analysis:
\begin{equation}
	\hat{H}=\frac{\hat{P}_x^2}{2m}+\frac{1}{2}m\omega^2\hat{X}^2+\frac{1}{2}[f(\hat{T})\hat{X}+\hat{X}f(\hat{T})]+g(\hat{T})\hat{P}_x\label{e7}
\end{equation}
The corresponding effective commutative Schr\"odinger equation can be obtained by taking overlap of (\ref{e74}) in coherent state basis (\ref{A8},\ref{e45}),
\small
\begin{equation}
	i\partial_t \psi_{phy}(x,t)=\left[\frac{P_x^2}{2m}+\frac{1}{2}m\omega^2 X_{\theta}^2+\frac{1}{2}\{f(T_{\theta})X_{\theta}+X_{\theta}f(T_{\theta})\}+g(T_{\theta})P_x\right]\psi_{phy}(x,t)\label{e8}
\end{equation}
\normalsize
At this stage, it will be interesting to note that $X_{\theta}$ and $T_{\theta}$ can be related to commutative $x$ and $t$, defined in (\ref{47},\ref{48}), by making use of similarity transformations,
\begin{equation}
	X_{\theta}= SxS^{-1},~T_{\theta}=S^{\dagger}t(S^{\dagger})^{-1};~~S=e^{\frac{\theta}{4}(\partial_t^2+\partial_x^2)}e^{-\frac{i\theta}{2}\partial_t\partial_x}\label{M5}
\end{equation}  
This $S$, a non-unitary operator, can be used to define the following map,
\begin{equation}
	S^{-1}\,:\,\,L_*^2(\mathbb{R}^1)\rightarrow L^2(\mathbb{R}^1) ~;~~i.e.~ S^{-1}\big(\psi_{phy}(x,t)\big):= \, \psi_c(x,t) \in L^2(\mathbb{R}^1) \label{M6}
\end{equation}
Now one can easily verify at this stage,
\begin{equation}
	\Big \langle \psi_{phy}\, ,\, \phi_{phy} \Big\rangle_{*,\,t}= \langle\psi_c\,\,,\phi_c\rangle _t\,\,\,\,\forall\,\psi_{phy},\phi_{phy}\,\in\, L^2_*(\mathbb{R}^1)\label{M7}
\end{equation}
Thus, one can replace non-local Voros star product with the local point-wise multiplication as in the usual commutative QM but \textit{only} within the integral when working in $L^2(\mathbb{R}^1)$. This is because one can establish the identity only by dropping many of the boundary terms indicating that $\psi_{phy}^*(x,t)\star_{V}\psi_{phy}(x,t)\neq |\psi_c(x,t)|^2$. Thus here, one can't interpret $|\psi_c(x,t)|^2$ as the probability density at point $x$ at time $t$, unlike $(\psi_{phy}^*\star_{V}\psi_{phy})(x,t)$. In this context, one can recall the so-called T-map connecting Moyal and Voros star products \cite{bal11}.
Now, using (\ref{M5},\ref{M6}) in (\ref{e8}) and retaining terms upto linear in $\theta$, finally (\ref{e8}) can now be recast as,
\begin{equation}
	i\partial_t \psi_c(x,t)=H_c\psi_c(x,t)\label{e13}
\end{equation}
where the corresponding effective commutative Hamiltonian $H_c$ is given by
\small
\begin{equation}
	H_c= \alpha(t) p_x^2+\beta x^2+\gamma(t)(xp_x+p_xx)+f(t)x+g(t)p_x=H_{GHO}+ f(t)x+g(t)p_x\label{H}
\end{equation}
\normalsize
Here $H_{GHO}$ stands for the Hamiltonian of a generalised time-dependent harmonic oscillator representing the first three terms. The last two terms represent perturbations linear in position and momentum in coordinate basis. Finally, all the various coefficients in (\ref{H}) are given by,
\begin{equation}
	\alpha(t)=\frac{1}{2m}-\theta \dot{g}(t);\,\,\,\,
	\beta= \frac{1}{2}m\omega^2;\,\,\,\,
	\gamma(t)=-\frac{1}{2}\theta\dot{f}(t)\label{e14}
\end{equation}
This Hamiltonian can be put into the diagonal form-
\begin{equation}
	\tilde{H}_c=\Omega(t)(a^{\dagger}a+\frac{1}{2})= H_{GHO}\label{H4}
\end{equation}
after a series of time-dependent unitary transformations. For details we refer to \cite{partha2}. Carrying out the analysis in the Heisenberg picture with adiabatic approximation, we find after evolution through a cycle of time period $t=\mathcal{T}$,
\begin{equation}
	a^{\dagger}(\mathcal{T})=a^{\dagger}(0)exp \left[i \int _0^{\mathcal{T}} \Omega\,\, d\tau + i\int_0^{\mathcal{T}} \left(\frac{1}{\Omega}\right)\frac{d\gamma}{d\tau} d\tau \right]\label{e35}
\end{equation}
The second term in the exponential of (\ref{e35}) represents the additional phase  over and above the dynamical phase $e^{i\int \Omega(t)dt}$. This extra phase can now be written in a more familiar form of Berry phase \cite{berry}, given as a functional of the closed loop $\Gamma$,
\begin{equation}
	\Phi_G[\Gamma] =\oint_{\Gamma=\partial S} \frac{1}{\Omega} \nabla_\textbf{R}\gamma . d\textbf{R}=-\frac{\theta}{2}\int\int_S \nabla_{\textbf{R}}\left(\frac{1}{\Omega}\right) \times \nabla_{\textbf{R}} \Big(\dot{f}(t)\Big)\, .\, d\textbf{S}
	\label{e36}
\end{equation}
It is crucial to take note of the $su(1,1)$ Lie-algebraic structure of the GHO part of the effective Hamiltonian (\ref{H}) in noncommutative space-time that is responsible for this emergent geometrical phase.\\

\textbf{\underline{Acknowledgements:}}
\small
We sincerely thank Prof. V.P. Nair and Prof. T.R. Govindarajan for giving us the opportunity to contribute to this celebratory Festschrift to honour the legacy of Prof. A.P. Balachandran in his 85th birthday.

\end{document}